\documentclass[11pt,prd,onecolumn,showpacs,amsmath,amssymb,aps,floats,floatfix]{article}
\usepackage{aas_macros}
\usepackage{placeins}
\usepackage[normalem]{ulem}
\usepackage[flushleft]{threeparttable}
\usepackage[sort,numbers]{natbib}
\usepackage{authblk}
\usepackage{gensymb}
\usepackage{hyperref}
\usepackage{comment}
\usepackage{dcolumn}
\usepackage{bm}
\usepackage{amsmath}
\usepackage{amsfonts}
\usepackage{amssymb}
\usepackage{color}
\usepackage{latexsym}
\usepackage{slashed} % slash mark
\usepackage{indentfirst}
\usepackage{mathrsfs}
\usepackage{multirow}
\usepackage{setspace} % spacing

\usepackage{amsmath}
\usepackage{float}
\usepackage{footnote}

\usepackage{indentfirst}

\usepackage{graphicx}
\usepackage{mathrsfs}

\usepackage{slashed} % slash mark

\def \be {\begin{equation}}
\def \ee {\end{equation}}
\def \ba {\begin{eqnarray}}
\def \ea {\end{eqnarray}}
\def\GeV{\mathrm{GeV}}     % GeV
\newcommand {\dbar}{{d\kern-.22em\lower-.73ex\hbox{-}}}

\newcommand {\fslash}[1]{{#1\kern -0.6em / \kern 0.2em}}

\def\s{\mathrm{s}}
\def\AU{\mathrm{AU}}

\def\sr{\mathrm{sr}}

\def\GV{\mathrm{GV}}

\def\cm{\mathrm{cm}} % cm
\def\GeV{\mathrm{GeV}} % GeV
\def\m{\mathrm{m}}

\makeindex

\begin{document}
\title{Solar modulation of cosmic proton and helium with AMS-02}
\author[1,2]{Bing-Bing Wang}
\author[3,4]{Xiao-Jun Bi}
\author[3]{Kun Fang}
\author[5]{Sujie Lin}
\author[3]{Peng-Fei Yin}
\affil[1]{Department of Space Science, University of Alabama in Huntsville, Huntsville, AL 35899, USA}
\affil[2]{Center for Space Plasma and Aeronomic Research (CSPAR), University of Alabama in Huntsville, Huntsville, AL 35899, USA}
\affil[3]{Key Laboratory of Particle Astrophysics, Institute of High Energy Physics,
Chinese Academy of Sciences, Beijing 100049, China}
\affil[4]{School of Physical Sciences, University of Chinese
Academy of Sciences, Beijing 100049, China}
\affil[5]{School of Physics and Astronomy, Sun Yat-Sen University, Zhuhai 519082, GuangDong, China}
\maketitle

%\begin{landscape}
%\begin{minipage}{1\columnwidth}%
%\begin{minipage}{1\columnwith}%

\begin{abstract}
We investigate the solar modulation effect with the long time cosmic ray proton and helium spectrum measured by AMS-02 on the time scale of a Bartels rotation (27 days) between May 2011 and May 2017.
The time-span covers the negative heliospheric magnetic field polarity cycle, the polarity reversal period and the positive polarity cycle.
The unprecedented accuracy of AMS-02 observation data provide a good opportunity to improve the understanding of the time dependent solar modulation effect.
In this work, a two-dimensional solar modulation model is used to compute the propagation of cosmic rays in the heliosphere.
Some important ingredients of the model which reflect the global heliospherical 
environment
% such as the solar wind speed, the heliospheric magnetic field strength, and the 
% tilt angle of the heliospheric current sheet,
are taken from the observations.
The propagation equation is numerically solved with the pubic Solarprop code.
We find that the drift effect is suppressed during the high solar activity period but nearly recovered in the first half of 2017.
The time-dependent rigidity dependence of the mean free path is critical to reproduce the observations between August 2012 and October 2015.
We also confirm that the proton and helium have the same diffusive mean free path.
The future monthly AMS-02 and PAMELA data will further confirm the vital assumption on the universal mean free path for all species; the antiproton data will be crucial to determine the drift effect during different epochs.
\end{abstract}

\section{Introduction}
The galactic cosmic rays are believed mainly come from supernova remnants. After injected from the sources, cosmic rays propagate in the interstellar space.
When cosmic rays enter the heliosphere, the interaction with the solar wind and the embedded magnetic field results in the intensity and the spectral shape of low energy cosmic rays are different from the local interstellar spectrum (LIS) \cite{1971RvGSP...9...27J,2013LRSP...10....3P}.
These effects on cosmic rays are called solar modulation.
The solar modulation effect limits our understanding for cosmic rays outside the heliosphere.
Therefore, the study of solar modulation is important for studying the injection and propagation parameters of cosmic rays, dark matter indirect measurement and the diffusion theory in the galaxy and heliosphere \cite{2017ApJ...840..115B,2017arXiv171203178T,2018ApJ...858...61B,2018ApJ...854...94B,2018ApJ...863..119Z,1973ApJ...182..585J,2002A&A...393..703T,2003AdSpR..32..549B,2004GeoRL..3110805B,2004JGRA..109.1109P,2009ASSL..362.....S,2010JGRA..115.3103P,2018ApJ...856...94Z,2018ApJ...854..137S}.

The recent experimental results from Voyager 1, PAMELA, and AMS-02 have achieved great breakthroughs which are useful to understand the solar modulation effect.
The Voyager 1 flew outside the heliosphere on August 2012 and directly measured the LIS in the range from a few to hundreds MeV/nucleon \cite{2013Sci...341..150S,2013Sci...341.1489G,2016ApJ...831...18C}.
The monthly PAMELA measurements of proton spectra \cite{2013ApJ...765...91A,2018ApJ...854L...2M} shed light on some details of the solar modulation \cite{2014SoPh..289..391P,2015ApJ...815..119V,2016ApJ...829....8C,2016AdSpR..57.1965R,2017ApJ...849L..32T,2017ApJ...846...56Q,2019PhRvD.100f3006W}.
Recently, the AMS-02 collaboration has published the continuous proton and helium energy spectrum with rigidity above $1\,\GV$ for proton and $1.9\,\GV$ for helium between May 2011 and May 2017 \cite{Aguilar:2018wmi}.
Some important results have been obtained, such as the confirmation of the velocity dependence of cosmic ray diffusion and finding the increase of the slope of perpendicular mean free path during solar maximum for low rigidity particles \cite{2018PhRvL.121y1104T,2019ApJ...871..253C}.

In \cite{2019PhRvD.100f3006W} (hereafter Paper I), we build a modulation model 
to well reproduce the long time PAMELA proton measurements between July 2006 to 
February 2014.
%In Paper I, we found that the drift effect is suppressed and the rigidity dependence of the mean free path changes over time during the polarity reversal period (assumed between November 2012 and March 2014).
This modulation model includes mainly physical processes affecting the 
propagation of cosmic rays in the heliosphere: diffusion, convection, drift, and 
energy loss.
Meanwhile, some main factors affecting solar modulation are taken from 
observations, such as the magnitude of the heliospheric magnetic field, the 
solar wind speed, and the tilt angle of the heliospheric current sheet (HCS).
We adopt the same model in this work.
We also deliberate to keep the model as simple as possible by only including the 
minimal degree of freedom.
The 2D modulation code solarprop 
\footnote{\url{http://www.th.physik.uni-bonn.de/nilles/people/kappl/}}\cite{
2016CoPhC.207..386K} is used to solve the cosmic ray propagation equation and 
obtain the modulated spectra.
In Paper I, we find that
the modulation processes are different between negative and polarity reversal 
period.
Therefore, we analyze the modulation effect with AMS-02 proton data separately 
in three periods related to the magnetic field polarity.
After successfully reproducing the proton observations, the modulation 
parameters for proton are applied to calculate the modulated helium spectrum.

The paper is organized as follows. In Section ~\ref{sec:model}, we briefly 
introduce the modulation model and present the LIS for proton and helium.
In Section ~\ref{sec:proton}, we compute the modulated proton spectrum and 
compare them to observations.
In Section ~\ref{sec:helium}, we check the assumption that proton and helium 
have the same mean free path by computing the helium spectrum with the 
modulation parameters for proton.
Finally, we give a summary in Section ~\ref{sec:conclusion}.

\section{A 2D solar modulation model}\label{sec:model}
There are four major modulation mechanisms for cosmic rays in the heliosphere:
diffusion on irregularities of the heliospheric magnetic field, convection by 
the outward solar wind, particles drift in the non-uniform magnetic field and 
adiabatic energy loss.
Several review articles discuss the modulation process in great detail 
\cite{1971RvGSP...9...27J,2013LRSP...10....3P}.
The propagation process can be described by the Parker equation 
\cite{1965P&SS...13....9P}:
\begin{equation}
\frac{\partial f}{\partial t} = -(\vec{V}_{sw} +\vec{V}_{drift})\cdot \nabla f + 
\nabla \cdot[K^s\cdot \nabla f] + \frac{\nabla \cdot \vec{V}_{sw}}{3} 
\frac{\partial{f}}{{\partial\ln p}} ,
\end{equation}
where $f(\vec{r},p,t)$ is the omni-directional distribution function, $\vec{r}$ 
is the position in the heliocentric spherical coordinate system, $p$ is the
particle momentum, $\vec{V}_{sw}$ is the solar wind speed, $\vec{V}_{drift}$ is 
the drift speed, $K^s$ is the symmetric part of diffusion tensor. The 
differential intensity related with the distribution function is given by $I = 
p^2f$.

It is customary to assume that the diffusion coefficient can be separated into 
spatial and rigidity components \cite{2013LRSP...10....3P}.
The generally assumption about the rigidity part is that all particle species 
have a universal function of rigidity 
\cite{1971Ap&SS..11..288G,1971JGR....76..221F,1975JGR....80.1701F,
2013SSRv..176..299M,2018AdSpR..62.2859B,2017ICRC...35...24V}.
In Paper I, we adopt a linear rigidity dependence of the diffusion coefficient 
and can reproduce the PAMELA monthly proton measurements between 2006 to 2012.
But during the polarity reversal period which is assumed between November 2012 
and March 2014 in Paper I based on Ref. \cite{2015ApJ...798..114S}, the time-dependent rigidity dependence is necessary 
to reproduce the observations.
In the present work, the parallel diffusion coefficient is adopted as the 
following form taking into account the finding in Paper I:
\begin{equation}
    k_{\parallel} =  \frac{1}{3} k \beta (\frac{R}{1\,\GV})^{\delta} 
\frac{B_E}{B} \end{equation}
where $k = 3.6 \times 10^{22}k_0~\cm^2/\s$ is a scale factor to model the time 
dependence of the diffusion coefficient, $\beta$ is the particle speed in the 
unit of the speed of light,
$\delta$ determines the rigidity dependence of the diffusion coefficient  (by 
default $\delta =1$),
$B_E$ is the heliospheric magnetic field strength near the Earth, 
$B=B_0/{r^2}\sqrt{1+\tan{\psi}}$ is the heliospheric magnetic field strength at 
the particle position and $\psi$ is the angle between magnetic field direction 
and its radial direction \cite{1958ApJ...128..664P}.
The standard Parker magnetic field model \cite{1958ApJ...128..664P} is used in 
this work.
We take the perpendicular diffusion coefficient to be $k_{\perp} = 0.02 
k_{\parallel}$ according to the test particle simulation 
\cite{1999ApJ...520..204G}.
The diffusion coefficient is also often marked as $k_{\parallel/\perp}=\frac{1}{3}v \lambda_{\parallel/\perp}$, 
where $v$ is particle speed and $\lambda_{\parallel}$ ($\lambda_{\perp}$) is called the parallel (perpendicular) mean free path.

The gradient and curvature drift speed is written as $V_{gc} = q \frac{\beta 
R}{3} \, \nabla \times \frac{\vec{B}}{B^2}$ \cite{1977ApJ...213..861J}.
We describe the heliospherical current sheet (HCS) drift following Ref.
\cite{1989ApJ...339..501B}, where a thick, symmetric transition region 
determined by the tilt angle is used to simulate a wavy neutral sheet.
The HCS drift speed $V_{ns}^{w}$ is given by
\begin{equation}
   \begin{aligned}
		\vec{V}_{ns}^{w}  = & \begin{cases}
		qA \frac{v \theta_{\triangle} 
\cos(\alpha)}{6\sin(\alpha+\theta_{\triangle})} \vec{e}_{r} ,& 
\pi/2-\alpha-\theta_{\triangle} < \theta <\pi/2 + \alpha +\theta_{\triangle} \\
      0, & else
	\end{cases}
   \end{aligned}
\end{equation}
where $q$ is the charge sign and $v$ is the particle speed, $\theta_{\triangle} 
\approx \frac{2RV_{sw}}{B_0\Omega \cos{\alpha}}$.
When the polar solar magnetic field directs outward in the north (southern) 
hemisphere and inward in the southern (north) hemisphere, it is said that the 
Sun is in a positive (negative) polarity cycle marked as $A>0$ ($A<0$).
$qA$ determines the drift direction.
Taking into account the possible suppression of the drift effect, a scale factor 
$k_d$ (by default $k_d=1$) is introduced and the drift velocity is described as 
$V_{drift} = k_d(V_{gc}+V_{ns}^{w})$ 
\cite{2003AdSpR..32..645F,2015ApJ...810..141P}.

The solar wind speed and the magnitude of magnetic field are taken from the 
website {\url{omniweb.gsfc.nasa.gov}}.
The tilt angle of HCS is obtained from the website \url{wso.stanford.edu} with 
the ``new'' model.%which is more accurate than the ``classic'' model.
These quantities are averaged over several months which corresponds to the time 
of solar wind propagation from the Sun to the modulation boundary at $100 
\,\AU$.
More elaborate description to our model is given in Paper I and references 
therein.
The discussion about the modulation resulted from the merged interaction regions 
\cite{1993ApJ...403..760P,1993AdSpR..13..239P,2019ApJ...878....6L} is not 
included in the recent work.

As an initial input condition in the modulation model, the LIS are constrained 
by the current experimental measurements.
Voyager 1 has directly measured the LIS in the range from a few to hundreds of
MeV/nucleon.
The monthly precise AMS-02 data provide important ingredients to reconstruct the 
LIS.
The LIS for proton and helium are constructed by the cubic spline interpolation 
method following the works \cite{2016A&A...591A..94G,2018ApJ...863..119Z,2019PhRvD.100f3006W}.
This method avoids the bias comes from the cosmic ray injection and propagation model which is still in debate.
We determine the proton and helium LIS by matching the low energy LIS to the 
Voyager 1 measurements and fitting the calculated spectrum to the AMS-02 data 
observed during Bartels rotation 2429, 2432, 2435 and 2438 
\cite{2016ApJ...831...18C,Aguilar:2018wmi}. These time periods are all within 
the negative polarity and the data can be well explained with one free 
parameter $k_0$.
The GNU Scientific Library 
(GSL)\footnote{\url{https://www.gnu.org/software/gsl/}} is used to perform the 
least-squares fitting.
The energy knots and the corresponding intensities in the cubic spline 
interpolation method are shown in Table \ref{tab:plis} and \ref{tab:helis}.
The difference between the LIS obtained by this work and that derived in Paper I 
is very small (see Appendix \ref{sec:liss}). 
Once the LIS have been derived, the effects of solar modulation are calculated 
directly from the model.

\begin{table}[!htbp]
   \caption{The parameterization of proton LIS with the cubic spline 
interpolation method. $E_k$ is kinetic energy and $I$ is intensity.}   \label{tab:plis}
   \centering
   \begin{tabular}{lcccccccl}\hline \hline
       $\mathrm{log(E_k/\GeV)}$            &-2.42    &-1.41  &-0.50   &0    
&0.50    &1.00    &1.50    &2.00 \\
       $\mathrm{log(I/(\GeV\m^2\sr\,\s))}$ &4.2905   &4.4688 &4.0176  &3.4548  
&2.5849  &1.4158  &0.0597  &-1.3497    \\
   \hline
   \end{tabular}
\end{table}

\begin{table}[!htbp]
   \caption{The parameterization of helium LIS with the cubic spline 
interpolation method. $E_k$ is kinetic energy and $I$ is intensity.}   \label{tab:helis}
   \centering
	\begin{tabular}{lccccccl}\hline \hline
      %&$log(R/\GV)$                    &-1.0        &-0.50     &0       &0.50    &1.00    &1.50       &2.00 \\
        $\mathrm{log(E_k/\GeV)}$                  &-2.27        &-1.28     &-0.30 
  &0.56    &1.22    &1.78       &2.29 \\
        $\mathrm{log(I/(\GeV\m^2\s\,\sr))}$   &2.3812      &2.7324 	 &2.6735 
 &1.7705  &0.4619 &-0.8980   &-2.2892 \\ \hline
   \end{tabular}
\end{table}

\section{Solar modulation for proton}\label{sec:proton}
The AMS-02 data are taken during different solar activity levels and different 
magnetic field conditions: the negative polarity cycle, the undefined polarity 
period around the solar maximum and finally the positive polarity cycle.
In Paper I, we show that the diffusion and the drift are both different during 
the negative and the polarity reversal periods.
Thus, we investigate the modulation effect for proton separately in three 
periods related to the magnetic field polarity.

\subsection{Modulation of CR proton with the assumption of the negative 
polarity} \label{sec:negative}
During every solar maximum the polarity of the solar magnetic field and 
subsequently the heliospherical magnetic field reverses direction.
After the polarity reversal took place around 2000 \cite{2003ApJ...598L..63G}, 
the polarity is negative until the recent reversal.
During the negative polarity cycle ($A<0$), positively charged particles drift into the 
inner heliosphere along the HCS and out over the poles.
Due to the asymmetric solar activity, the reversal of the solar magnetic field 
polarity is not simultaneous in two hemispheres.
The summary of some estimates of the solar polar field reversal times for the northern 
and southern solar hemisphere is presented in Table \ref{tab:time}.
\begin{table}[!htbp]
   \caption{Estimates of the time of solar polar magnetic field polarity 
reversals in the northern and southern hemisphere.}  \label{tab:time}
   \centering
	\begin{tabular}{ccc}\hline \hline
      North       	   &South      &Ref. \\
		2012/06	  			&-      	    		
&\cite{2014SoPh..289.3381K}     \\
		2012/11	  			&2014/03     		
&\cite{2015ApJ...798..114S}     \\
		2013/07	  			&2015/01		 		
&\cite{2015GeAe..55..969T} \\
		2012/05-2014/04	&2013/06-2015/03	
&\cite{2016KPCB...32...78P} \\
		2012/10-2015/09	&2014/06		 		
&\cite{2016ApJ...823L..15G}	\\
		2012/06-2014/11	&2013/10		 		
&\cite{2018AA...618A.148J}	\\
   \hline
\end{tabular}
\end{table}
One can see from Table \ref{tab:time} that the estimated reversal time can be 
very different by means of different methods and data.
The polarity reversal period is not well determined.
However, we also see that the polarity reversal process had not happened before 
the early of 2012. Thus, it is safe to set the heliospheric magnetic field polarity as negative in this 
time period.

\begin{figure}[hbt!]
   \centering
   \includegraphics[width=1\textwidth]{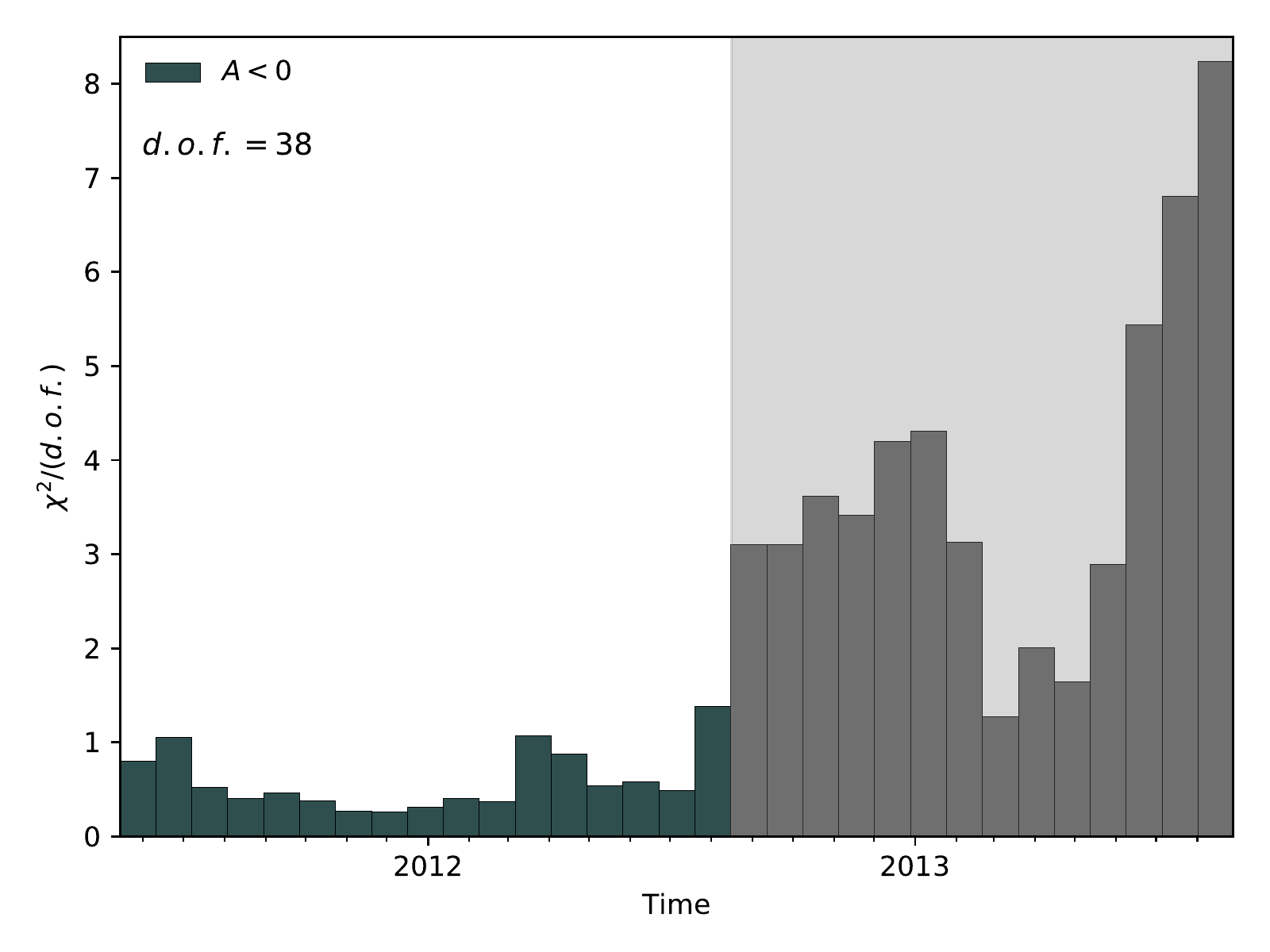}
   \caption{The time profile of reduced-$\chi^2$ for the fit to the monthly 
AMS-02 proton data between May 2011 and November 2012 under the assumption 
$A<0$. The scale factor of diffusion coefficient $k_0$ is taken as the free 
parameter.}
	\label{fg:am_chi2}
\end{figure}

We compare the computed spectrum with the AMS-02 measurements with rigidity 
below $40\,\GV$.
For the particle with higher rigidity, the modulation effect is negligible.
Following the scenario in Paper I which reproduces 6 years PAMELA proton 
spectrum between 2006 to 2012, we fixed both $k_d$ and $\delta$ as 1 and 
only adjust $k_0$ to fit the observation.
The resulting time profile of the reduced-$\chi^2$ ($\chi^2/(d.o.f.)$) is shown 
in Figure \ref{fg:am_chi2}.
The reduced-$\chi^2$ is around or less than 1 until August 2012.
There is no significant need to introduce more free parameter for this period.
However, after the August 2012, the reduced-$\chi^2$ suddenly increases to an 
unacceptable level with reduced-$\chi^2>2$. The model with only one free 
parameter $k_0$ fails to correctly describe the modulation process after this 
time node.

\subsection{Modulation of CR proton with the assumption of the positive 
polarity} \label{sec:positive}
As showed in Table \ref{tab:time}, the estimated latest time of the completion 
of the reversal is September 2015.
After the reversal, the solar magnetic field polarity becomes positive ($A>0$).
Positively charged particles drift into the inner heliosphere over the poles and 
out of it along the HCS. They have less difficulty in reaching the Earth and 
less modulation than that during the negative polarity period.

When the solar activity indicated by the sunspot number decreases to moderate %see from tilt angle paper I figure 1
level, we expect the turbulence magnetic properties and the rigidity 
dependence of the mean free path recover to the similar behavior with  
$\delta=1$ as in the negative polarity cycle.
We attempted to adopt the full drift effect ($k_d=1$)
, but the required scale factor of diffusion coefficient $k_0$ 
is much smaller than 0.7 prior to October 2016 (see Appendix \ref{sec:full}). Under the simple framework of the 
force-field approximation, the modulation potential is inversely proportional to 
the diffusion coefficient characterized by $k_0$.
%The variation of solar modulation potential reconstructed from the neutron monitor count rate \cite{2017AdSpR..60..833G,2019JGRA..124.2367K} is within 2.1 times during May 2011 to May 2017. If the full drift effect is 
The ratio of solar modulation potential (also the diffusion coefficient) reconstructed from the neutron monitor count rate \cite{2017AdSpR..60..833G,2019JGRA..124.2367K} is within 2.1 during May 2011 to May 2017. If the full drift effect is 
adopted, the required diffusion coefficient is too small and far from this relation.
In addition, this scenario results in larger $\chi^2$ 
than the suppressed drift case (see Appendix \ref{sec:full}). Thus, we set 
$k_d$ as a free parameter in this time period to reduce the drift effect.

\begin{figure}[hbt!]
   \centering
   \includegraphics[width=1\textwidth]{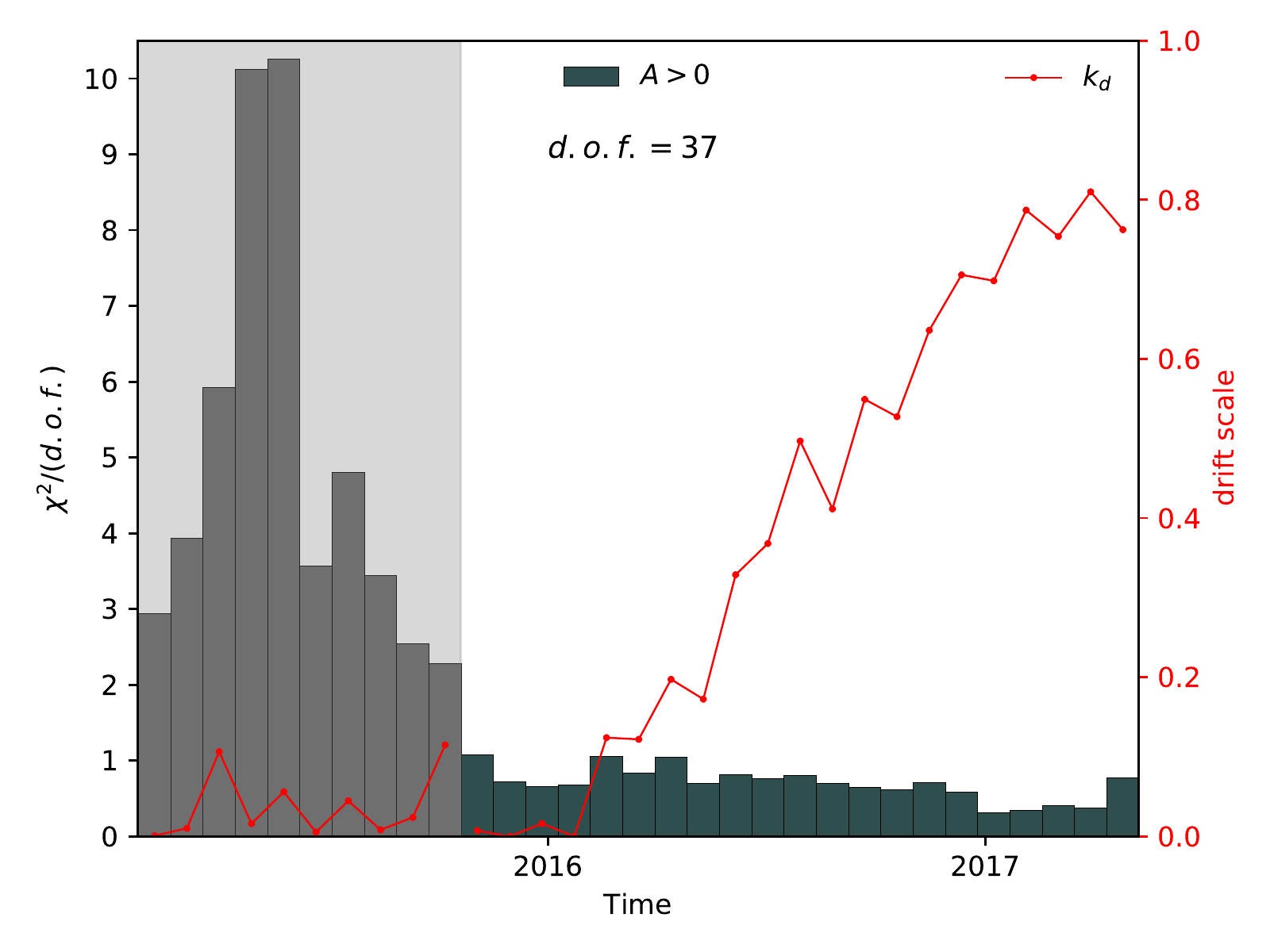}
  \caption{The time profile of reduced-$\chi^2$ and the drift scale factor $k_d$ from the fits to the 
AMS-02 proton data during August 2015 to May 2017 under the assumption of $A>0$. 
The scale factor for diffusion coefficient and drift velocity, $k_0$ and $k_d$ 
, are free parameters in this period.}
\label{fg:ap_chi2}
\end{figure}

In Figure \ref{fg:ap_chi2} we show that the time profile of reduced-$\chi^2$ and 
the scale factor for the drift speed $k_d$.
The reduced-$\chi^2$ values are less than or close to 1 during November 2015 to 
May 2017. The scale factor of drift velocity $k_d$ is nearly 0 around the 
second half of 2015, and it is about 0.8 in March 2017.
It indicates that the drift effect is suppressed during this period.
The increase of $k_d$ form 0 to 0.8 indicates the gradual recovery of the 
drift effect and implies that the drift effect may fully recover around the 
middle of 2017. 
There are several mechanisms that may cause the suppression of 
the drift effect. The large-scale fluctuations in the heliospheric magnetic 
field, such as the interaction regions and the merged interaction regions, fill 
the heliosphere so that drifts may only occur on a less scale during moderate 
to high solar activity period \cite{1993AdSpR..13..239P}.
In addition, numerical simulation shows that the presence of scattering can 
also suppress the drift effect. For an intermediate degree of scattering, the 
drift velocity is typically suppressed by a larger degree; when the scattering 
is very strong, there is no large-scale drift motions 
\cite{2007ApJ...670.1149M}.

Note that there is some degeneracy between $k_0$ and $k_d$ as shown in 
Appendix \ref{sec:deg}. As the drift effect is opposite for particle with 
opposite charge, a simultaneous fit to the proton and future antiproton 
spectrum is crucial to reduce the uncertainty and get a better understanding for 
the drift effect.

\subsection{Modulation for proton between August 2012 and October 
2015}\label{sec:ps}

From Section \ref{sec:negative}, we show that the default model with the linear 
rigidity dependence on the mean free path and the full drift effect ($\delta=1$, 
$k_d=1$) fails to describe the modulation process since August 2012.
Additionally, the model with linear rigidity dependence on the mean free path 
and the suppressed drift effect ($\delta=1$, $k_d \in [0,1]$) used in Section 
\ref{sec:positive} can not reproduce the observations before October 2015.
In Paper I, after we tested various configurations for the diffusion 
coefficient and drift effect to reproduce the PAMELA proton observations during 
November 2012 to February 2014
, we concluded that the combination of the time-dependent power-law rigidity 
dependence on the mean free path and the zero drift configuration give the best 
fit to the data.
So in this case, the free parameters are $k_0$ and $\delta$.
This scenario is adopted to reproduce the AMS-02 observations between August 
2012 and October 2015 which coincides with some estimated polarity reversal periods.

\begin{figure}[hbt!]
    \centering
   \includegraphics[width=1\textwidth]{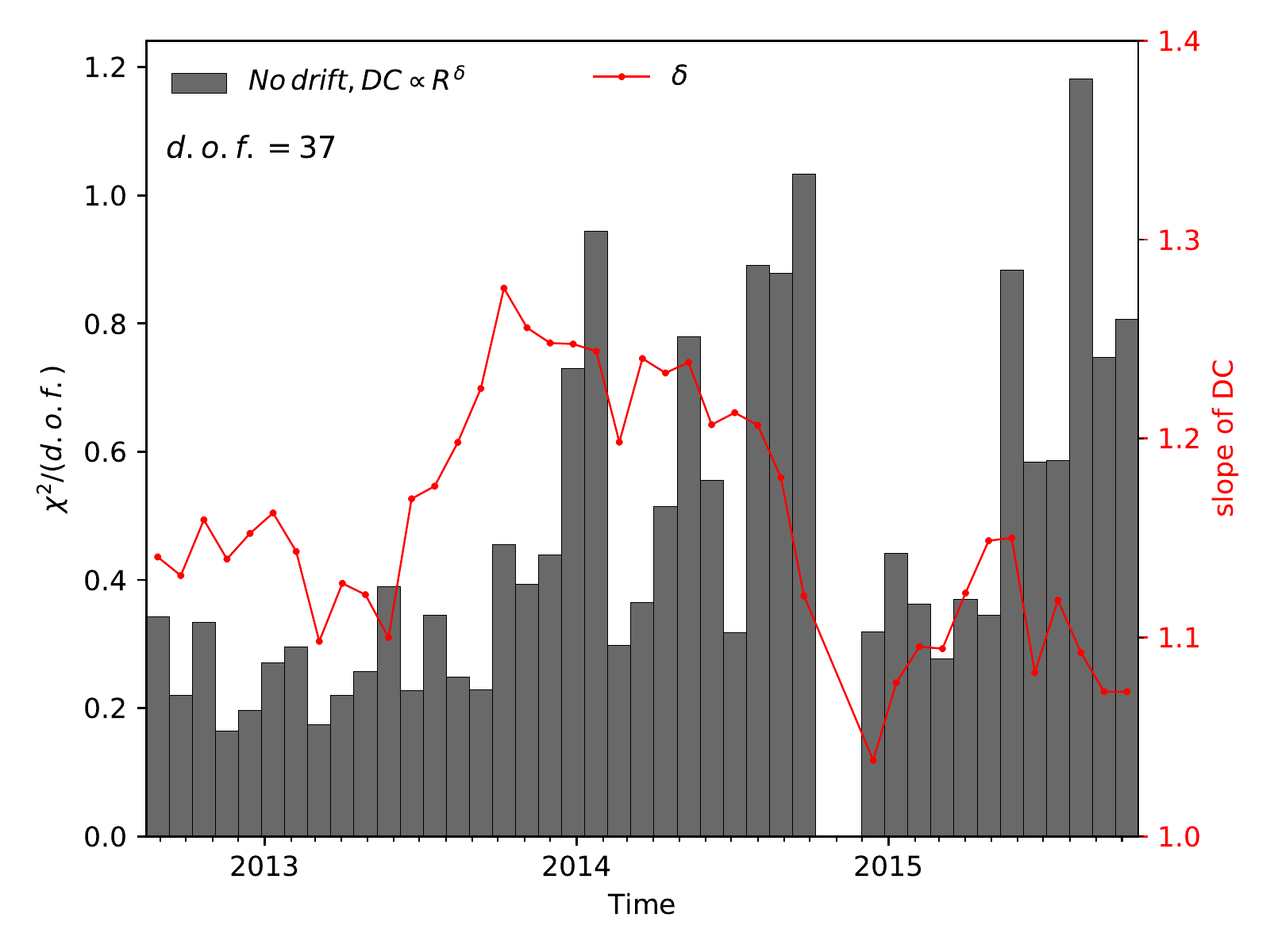}
    \caption{The time profile of reduced-$\chi^2$ and the slope of the mean free 
path $\delta$ for the fits to the AMS-02 proton data between August 2012 and 
October 2015. The scale factor for the diffusion coefficient $k_0$ and slope for 
the diffusion coefficient $\delta$ are free parameters in this period.}
\label{fg:ps_chi2}
\end{figure}

The time profile of the slope of the mean free path $\delta$ and the 
reduced-$\chi^2$ are shown in Figure \ref{fg:ps_chi2}.
We find that $\delta$ roughly keeps increasing until it reaches the maximum 
value of 1.28 in October 2013 and then decreases to 1.07 in October 2015.
The variation of rigidity dependence should be noticed in all cosmic ray 
species, such as helium.
Almost all the reduced-$\chi^2s$ are smaller than 1.
Obviously, the combination of the two parameters $k_0$ and $\delta$ is adequate 
to reproduce the observations.
Although introducing $k_d$ as the third free parameter may further improve the 
fit,
but the parameter space will not be constrained well because of the 
degeneracy between diffusion and drift parameters.
The future monthly antiproton data is needed to reduce the degeneracy.
The drift velocity is assumed to be 0 in this subsection, which may not be 
realistic in the whole period since it may have a transition process.
Because of the degeneracy between the scale factor ($k_0$), the slope ($\delta$) 
of the diffusion coefficient and the drift speed ($k_d$),
these transition processes have to be studied separately in greater detail.

\section{Solar modulation for helium}\label{sec:helium}
\begin{figure}[hbt!]
    \centering
   \includegraphics[width=1\textwidth]{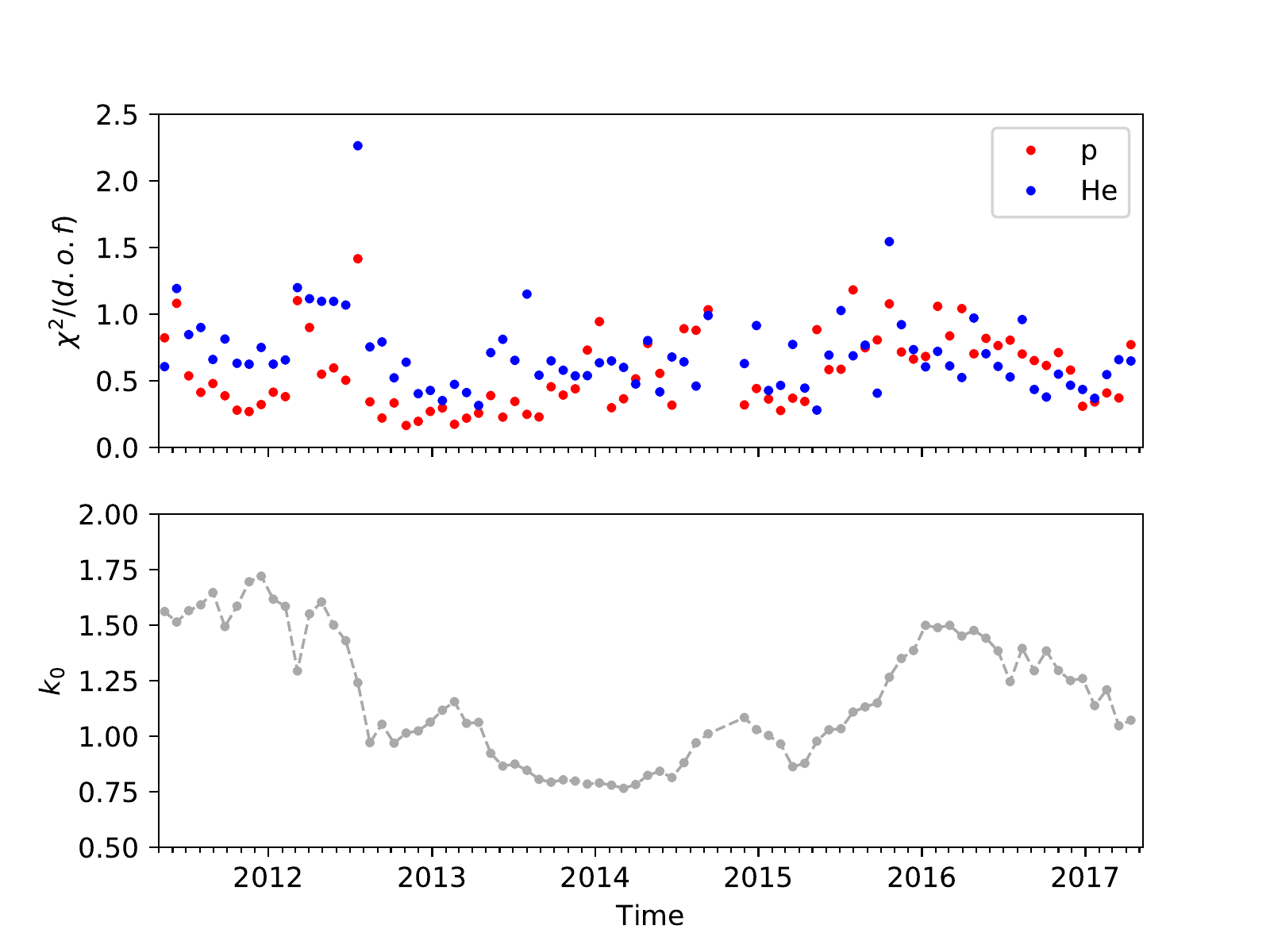}
    \caption{The upper panel shows the time profile of reduced-$\chi^2$ for 
proton (red dots) and helium (blue dots). The bottom panel shows the best fit 
$k_0$ for proton. Note that in the upper panel, the modulation parameters for 
helium are taken from proton.}
\label{fg:sum}
\end{figure}

\begin{figure}[hbt!]
     \centering
     \begin{tabular}{@{}cc@{}}
     \includegraphics[width=0.65\textwidth]{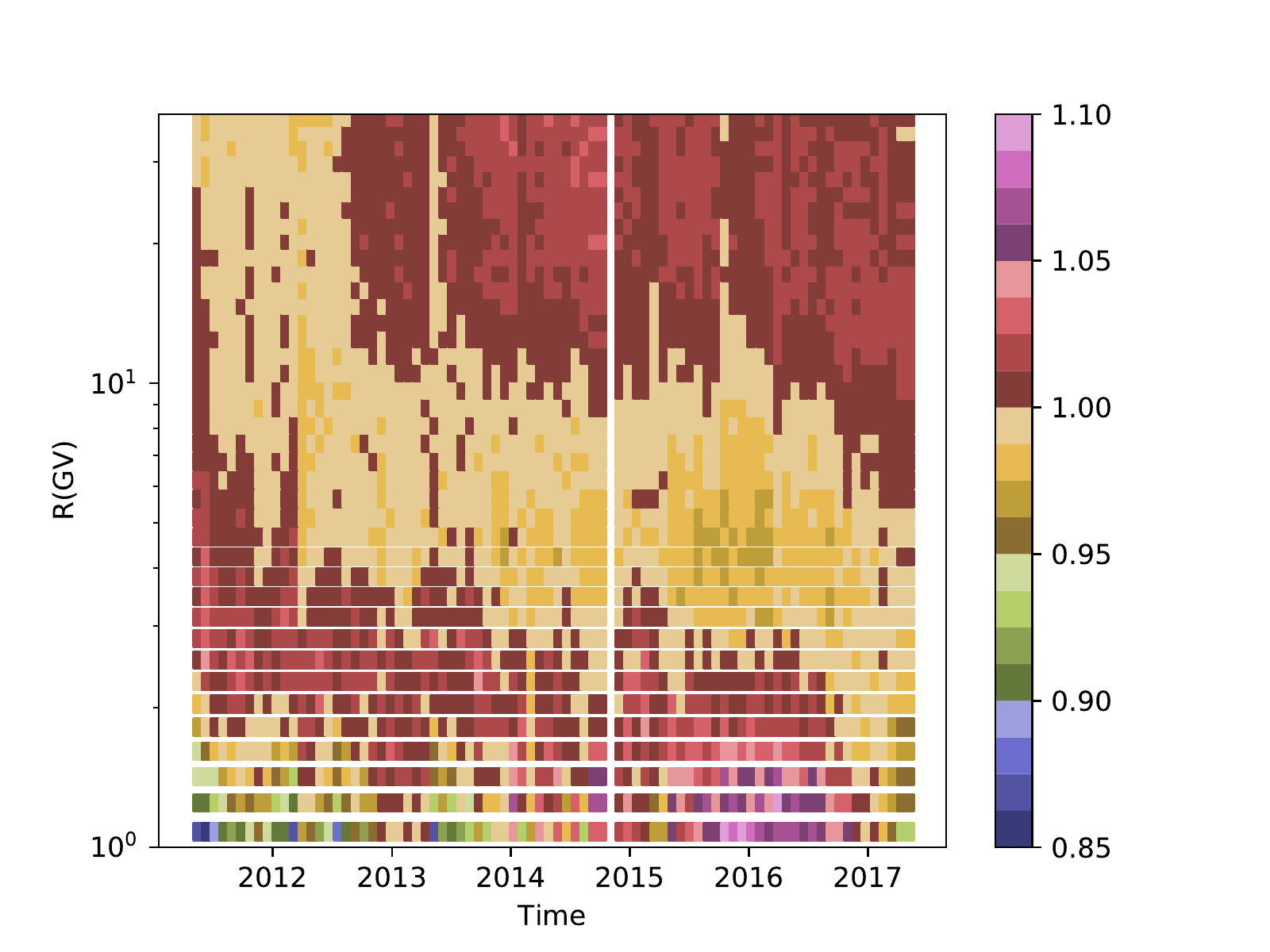}
     \includegraphics[width=0.65\textwidth]{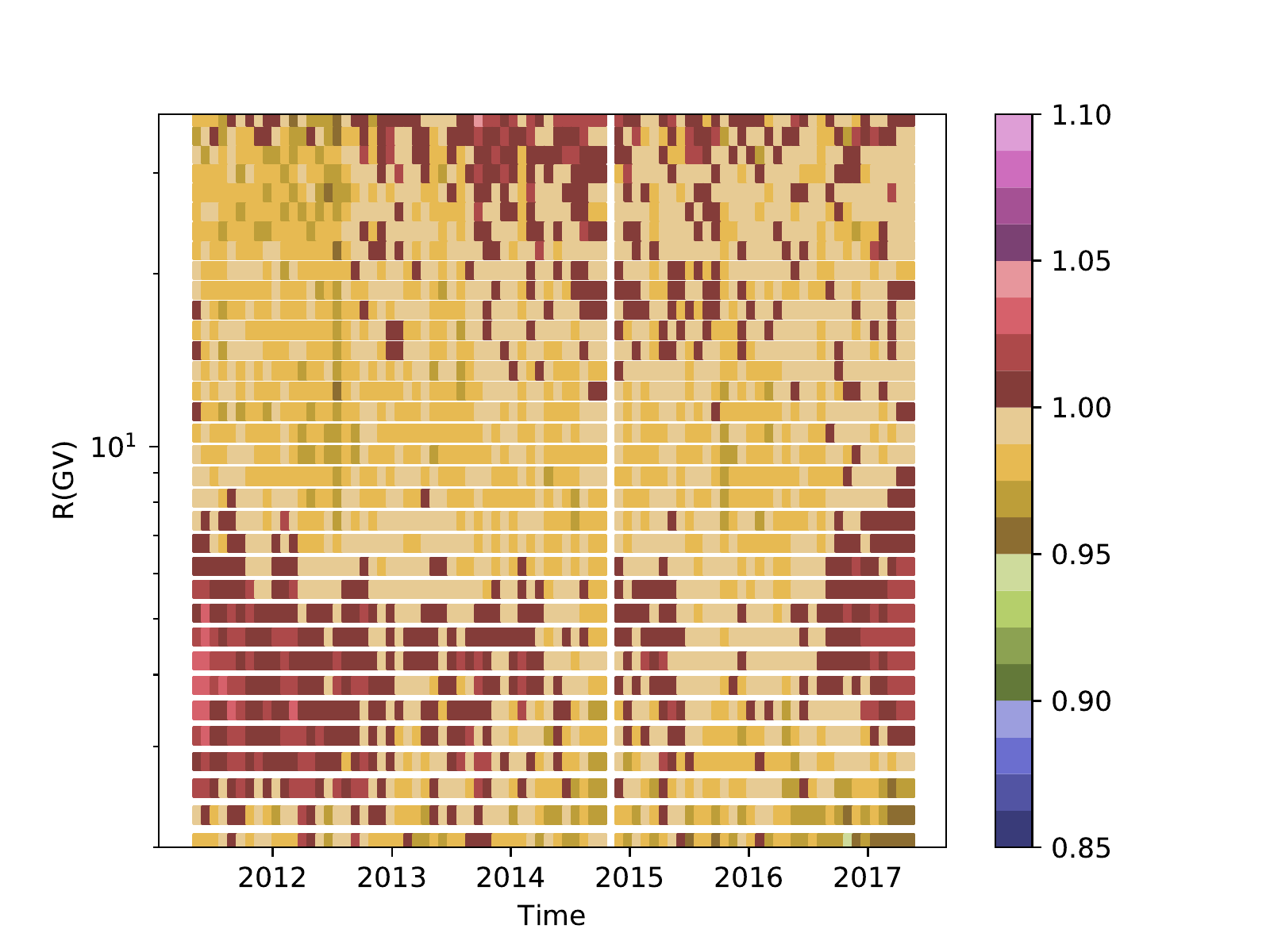}
     \end{tabular}
     \caption{The left (right) panel show the ratio of the computed proton 
(helium) intensities to measured values. The same modulation parameters are 
applied for proton and helium.}
     \label{fg:ratio}
\end{figure}

It is an important assumption that the mean free path is the same for all 
species of nuclei.
The precise AMS-02 measurements provide a good opportunity to check this widely 
adopted assumption. The main parameter $k_0$ for proton is shown in the bottom 
panel of Figure \ref{fg:sum}. The time profiles of $k_d$ and $\delta$ are 
shown in Figure \ref{fg:ap_chi2} and Figure \ref{fg:ps_chi2}, respectively.
We take the modulation parameters ($k_0$, $\delta$, $k_d$) obtained in the 
previous section as inputs to directly compute the modulated spectrum for 
helium.
The reduced-$\chi^2$ for helium and proton are summarized in the upper panel of 
Figure \ref{fg:sum}. We find that the same modulation parameters can 
well reproduce the proton and the helium observations simultaneously.
Meanwhile, we show the ratios of the computed intensities to the measured 
intensities as functions of rigidity and time in Figure \ref{fg:ratio}.
It can be seen that most of the fits agree with the data within $\pm5\%$.

Recalling the treatment to the modulation for boron and carbon in Paper I, 
in which the same mean free paths are able to reproduce the ACE boron and 
carbon observations, different nuclei could have 
the same mean free path in the heliosphere.
The upcoming monthly helium data of PAMELA between July 2006 and January 2016 
\cite{2019EPJWC.20901004M} and future time-dependent nuclei data of AMS-02 will 
give it a further test.

%\FloatBarrier
\section{Conclusion}\label{sec:conclusion}
The precise measurements of monthly cosmic ray proton and helium spectra by 
AMS-02 between May 2011 to May 2017 provide an important chance to improve our 
understanding for the solar modulation.
Compared to the PAMELA data up to February 2014, the AMS-02 data cover the whole 
solar magnetic field polarity reversal period around the solar maximum and part 
of the positive polarity cycle.
Meanwhile, the precise measurements of monthly helium spectrum provide a chance 
to check the important assumption that the cosmic ray proton and helium 
have the same diffusive mean free path in the 
heliosphere.

A two-dimensional model is used to describe the propagation of proton and helium 
in the heliosphere.
The model includes all major modulation processes and the variation of the 
heliosphere environment, such as the solar wind speed, the magnetic field 
strength and the tilt angle of HCS.
We consider a simplest reasonable scenario to reproduce the observations.
With no more than two free parameters, the computed spectrum are able to match 
the AMS-02 proton and helium observations.

We find that the rigidity dependence of the mean free path is varied with time. 
The linear rigidity dependence is adequate to reproduce the observations before 
August 2012 or after October 2015. 
Within the possible polarity reversal period between August 2012 and October 2015, 
the time varying power-law dependence of the mean free path is essential to fit the data. 
%The time varying power-law dependence of the mean free path is essential to fit the data between August 2012 and October 2015 which may suffer effects from the heliospheric polarity reversal process.  
We also find that the zero drift 
effect can well reproduce the observations during the polarity reversal 
period and the suppressed drift effect clearly keeps recovering after this 
period.
Finally, with the help of the precise monthly helium measurements, we confirm 
that the mean free path is the same for proton and helium.
The future monthly data from AMS-02 and PAMELA for other nuclei will provide further 
checking for the assumption that all nuclei have a universal mean free path,
and the monthly antiproton data would provide invaluable help to understand the 
role of the drift effect in different solar activity periods.

\section*{Acknowledgement}
This work is supported by the National Key R\&D Program of China (No. 
2016YFA0400200),
the National Natural Science Foundation of China (Nos. U1738209 and 11851303).

\bibliography{paper_ams}

\begin{thebibliography}{10}

\bibitem{1971RvGSP...9...27J}
J.~R. {Jokipii}.
\newblock {Propagation of cosmic rays in the solar wind.}
\newblock {\em Reviews of Geophysics and Space Physics}, 9:27--87, 1971.

\bibitem{2013LRSP...10....3P}
M.~S. {Potgieter}.
\newblock {Solar Modulation of Cosmic Rays}.
\newblock {\em Living Reviews in Solar Physics}, 10:3, June 2013.

\bibitem{2017ApJ...840..115B}
M.~J. {Boschini}, S.~{Della Torre}, M.~{Gervasi}, D.~{Grandi},
  G.~{J{\'o}hannesson}, M.~{Kachelriess}, G.~{La Vacca}, N.~{Masi}, I.~V.
  {Moskalenko}, E.~{Orlando}, S.~S. {Ostapchenko}, S.~{Pensotti}, T.~A.
  {Porter}, L.~{Quadrani}, P.~G. {Rancoita}, D.~{Rozza}, and M.~{Tacconi}.
\newblock {Solution of Heliospheric Propagation: Unveiling the Local
  Interstellar Spectra of Cosmic-ray Species}.
\newblock {\em \apj}, 840:115, May 2017.

\bibitem{2017arXiv171203178T}
Nicola {Tomassetti}.
\newblock {Solar Modulation of Galactic Cosmic Rays: Physics Challenges for
  AMS-02}.
\newblock {\em ArXiv e-prints}, page arXiv:1712.03178, December 2017.

\bibitem{2018ApJ...858...61B}
M.~J. {Boschini}, S.~{Della Torre}, M.~{Gervasi}, D.~{Grandi},
  G.~{J{\'o}hannesson}, G.~{La Vacca}, N.~{Masi}, I.~V. {Moskalenko},
  S.~{Pensotti}, T.~A. {Porter}, L.~{Quadrani}, P.~G. {Rancoita}, D.~{Rozza},
  and M.~{Tacconi}.
\newblock {Deciphering the Local Interstellar Spectra of Primary Cosmic-Ray
  Species with HELMOD}.
\newblock {\em \apj}, 858:61, May 2018.

\bibitem{2018ApJ...854...94B}
M.~J. {Boschini}, S.~{Della Torre}, M.~{Gervasi}, D.~{Grandi},
  G.~{J{\'o}hannesson}, G.~{La Vacca}, N.~{Masi}, I.~V. {Moskalenko},
  S.~{Pensotti}, T.~A. {Porter}, L.~{Quadrani}, P.~G. {Rancoita}, D.~{Rozza},
  and M.~{Tacconi}.
\newblock {HelMod in the Works: From Direct Observations to the Local
  Interstellar Spectrum of Cosmic-Ray Electrons}.
\newblock {\em \apj}, 854:94, February 2018.

\bibitem{2018ApJ...863..119Z}
Cheng-Rui {Zhu}, Qiang {Yuan}, and Da-Ming {Wei}.
\newblock {Studies on Cosmic-Ray Nuclei with Voyager, ACE, and AMS-02. I. Local
  Interstellar Spectra and Solar Modulation}.
\newblock {\em \apj}, 863:119, August 2018.

\bibitem{1973ApJ...182..585J}
J.~R. {Jokipii}.
\newblock {Radial Variation of Magnetic Fluctuations and the Cosmic-Ray
  Diffusion Tensor in the Solar Wind}.
\newblock {\em \apj}, 182:585--600, June 1973.

\bibitem{2002A&A...393..703T}
A.~{Teufel} and R.~{Schlickeiser}.
\newblock {Analytic calculation of the parallel mean free path of heliospheric
  cosmic rays. I. Dynamical magnetic slab turbulence and random sweeping slab
  turbulence}.
\newblock {\em \aap}, 393:703--715, October 2002.

\bibitem{2003AdSpR..32..549B}
J.~W. {Bieber}.
\newblock {Transport of charged particles in the heliosphere: Theory}.
\newblock {\em Advances in Space Research}, 32:549--560, August 2003.

\bibitem{2004GeoRL..3110805B}
J.~W. {Bieber}, W.~H. {Matthaeus}, A.~{Shalchi}, and G.~{Qin}.
\newblock {Nonlinear guiding center theory of perpendicular diffusion: General
  properties and comparison with observation}.
\newblock {\em \grl}, 31:L10805, May 2004.

\bibitem{2004JGRA..109.1109P}
S.~{Parhi}, J.~W. {Bieber}, W.~H. {Matthaeus}, and R.~A. {Burger}.
\newblock {Heliospheric solar wind turbulence model with implications for ab
  initio modulation of cosmic rays}.
\newblock {\em Journal of Geophysical Research (Space Physics)}, 109:A01109,
  January 2004.

\bibitem{2009ASSL..362.....S}
A.~{Shalchi}, editor.
\newblock {\em {Nonlinear Cosmic Ray Diffusion Theories}}, volume 362 of {\em
  Astrophysics and Space Science Library}, 2009.

\bibitem{2010JGRA..115.3103P}
C.~{Pei}, J.~W. {Bieber}, B.~{Breech}, R.~A. {Burger}, J.~{Clem}, and W.~H.
  {Matthaeus}.
\newblock {Cosmic ray diffusion tensor throughout the heliosphere}.
\newblock {\em Journal of Geophysical Research (Space Physics)}, 115:A03103,
  March 2010.

\bibitem{2018ApJ...856...94Z}
L.~L. {Zhao}, L.~{Adhikari}, G.~P. {Zank}, Q.~{Hu}, and X.~S. {Feng}.
\newblock {Influence of the Solar Cycle on Turbulence Properties and Cosmic-Ray
  Diffusion}.
\newblock {\em \apj}, 856:94, April 2018.

\bibitem{2018ApJ...854..137S}
Z.~N. {Shen} and G.~{Qin}.
\newblock {Modulation of Galactic Cosmic Rays in the Inner Heliosphere over
  Solar Cycles}.
\newblock {\em \apj}, 854:137, February 2018.

\bibitem{2013Sci...341..150S}
E.~C. {Stone}, A.~C. {Cummings}, F.~B. {McDonald}, B.~C. {Heikkila}, N.~{Lal},
  and W.~R. {Webber}.
\newblock {Voyager 1 Observes Low-Energy Galactic Cosmic Rays in a Region
  Depleted of Heliospheric Ions}.
\newblock {\em Science}, 341:150--153, July 2013.

\bibitem{2013Sci...341.1489G}
D.~A. {Gurnett}, W.~S. {Kurth}, L.~F. {Burlaga}, and N.~F. {Ness}.
\newblock {In Situ Observations of Interstellar Plasma with Voyager 1}.
\newblock {\em Science}, 341:1489--1492, September 2013.

\bibitem{2016ApJ...831...18C}
A.~C. {Cummings}, E.~C. {Stone}, B.~C. {Heikkila}, N.~{Lal}, W.~R. {Webber},
  G.~{J{\'o}hannesson}, I.~V. {Moskalenko}, E.~{Orlando}, and T.~A. {Porter}.
\newblock {Galactic Cosmic Rays in the Local Interstellar Medium: Voyager 1
  Observations and Model Results}.
\newblock {\em \apj}, 831:18, November 2016.

\bibitem{2013ApJ...765...91A}
O.~{Adriani} et~al.
\newblock {Time Dependence of the Proton Flux Measured by PAMELA during the
  2006 July-2009 December Solar Minimum}.
\newblock {\em \apj}, 765:91, March 2013.

\bibitem{2018ApJ...854L...2M}
M.~{Martucci} et~al.
\newblock {Proton Fluxes Measured by the PAMELA Experiment from the Minimum to
  the Maximum Solar Activity for Solar Cycle 24}.
\newblock {\em \apjl}, 854:L2, February 2018.

\bibitem{2014SoPh..289..391P}
M.~S. {Potgieter}, E.~E. {Vos}, M.~{Boezio}, N.~{De Simone}, V.~{Di Felice},
  and V.~{Formato}.
\newblock {Modulation of Galactic Protons in the Heliosphere During the Unusual
  Solar Minimum of 2006 to 2009}.
\newblock {\em \solphys}, 289:391--406, January 2014.

\bibitem{2015ApJ...815..119V}
Etienne~E. {Vos} and Marius~S. {Potgieter}.
\newblock {New Modeling of Galactic Proton Modulation during the Minimum of
  Solar Cycle 23/24}.
\newblock {\em \apj}, 815(2):119, Dec 2015.

\bibitem{2016ApJ...829....8C}
C.~{Corti}, V.~{Bindi}, C.~{Consolandi}, and K.~{Whitman}.
\newblock {Solar Modulation of the Local Interstellar Spectrum with Voyager 1,
  AMS-02, PAMELA, and BESS}.
\newblock {\em \apj}, 829(1):8, Sep 2016.

\bibitem{2016AdSpR..57.1965R}
J.~L. {Raath}, M.~S. {Potgieter}, R.~D. {Strauss}, and A.~{Kopp}.
\newblock {The effects of magnetic field modifications on the solar modulation
  of cosmic rays with a SDE-based model}.
\newblock {\em Advances in Space Research}, 57(9):1965--1977, May 2016.

\bibitem{2017ApJ...849L..32T}
N.~{Tomassetti}, M.~{Orcinha}, F.~{Bar{\~a}o}, and B.~{Bertucci}.
\newblock {Evidence for a Time Lag in Solar Modulation of Galactic Cosmic
  Rays}.
\newblock {\em \apjl}, 849:L32, November 2017.

\bibitem{2017ApJ...846...56Q}
G.~{Qin} and Z.-N. {Shen}.
\newblock {Modulation of Galactic Cosmic Rays in the Inner Heliosphere,
  Comparing with PAMELA Measurements}.
\newblock {\em \apj}, 846:56, September 2017.

\bibitem{Aguilar:2018wmi}
M.~Aguilar et~al.
\newblock {Observation of Fine Time Structures in the Cosmic Proton and Helium
  Fluxes with the Alpha Magnetic Spectrometer on the International Space
  Station}.
\newblock {\em Phys. Rev. Lett.}, 121(5):051101, 2018.

\bibitem{2018PhRvL.121y1104T}
N.~{Tomassetti}, F.~{Bar{\~a}o}, B.~{Bertucci}, E.~{Fiandrini}, J.~L.
  {Figueiredo}, J.~B. {Lousada}, and M.~{Orcinha}.
\newblock {Testing Diffusion of Cosmic Rays in the Heliosphere with Proton and
  Helium Data from AMS}.
\newblock {\em Physical Review Letters}, 121(25):251104, December 2018.

\bibitem{2019ApJ...871..253C}
C.~{Corti}, M.~S. {Potgieter}, V.~{Bindi}, C.~{Consolandi}, C.~{Light},
  M.~{Palermo}, and A.~{Popkow}.
\newblock {Numerical Modeling of Galactic Cosmic-Ray Proton and Helium Observed
  by AMS-02 during the Solar Maximum of Solar Cycle 24}.
\newblock {\em \apj}, 871:253, February 2019.

\bibitem{2019PhRvD.100f3006W}
Bing-Bing {Wang}, Xiao-Jun {Bi}, Kun {Fang}, Su-Jie {Lin}, and Peng-Fei {Yin}.
\newblock {Time-dependent solar modulation of cosmic rays from solar minimum to
  solar maximum}.
\newblock {\em \prd}, 100(6):063006, Sep 2019.

\bibitem{2016CoPhC.207..386K}
R.~{Kappl}.
\newblock {SOLARPROP: Charge-sign dependent solar modulation for everyone}.
\newblock {\em Computer Physics Communications}, 207:386--399, October 2016.

\bibitem{1965P&SS...13....9P}
E.~N. {Parker}.
\newblock {The passage of energetic charged particles through interplanetary
  space}.
\newblock {\em Planetary and Space Science}, 13:9--49, January 1965.

\bibitem{1971Ap&SS..11..288G}
L.~J. {Gleeson} and I.~H. {Urch}.
\newblock {Energy losses and modulation of galactic cosmic rays.}
\newblock {\em \apss}, 11:288--308, May 1971.

\bibitem{1971JGR....76..221F}
L.~A. {Fisk}.
\newblock {Solar modulation of galactic cosmic rays, 2}.
\newblock {\em \jgr}, 76:221, 1971.

\bibitem{1975JGR....80.1701F}
G.~J. {Fulks}.
\newblock {Solar modulation of galactic cosmic ray electrons, protons, and
  alphas}.
\newblock {\em \jgr}, 80:1701--1714, May 1975.

\bibitem{2013SSRv..176..299M}
H.~{Moraal}.
\newblock {Cosmic-Ray Modulation Equations}.
\newblock {\em \ssr}, 176:299--319, June 2013.

\bibitem{2018AdSpR..62.2859B}
M.~J. {Boschini}, S.~{Della Torre}, M.~{Gervasi}, G.~{La Vacca}, and P.~G.
  {Rancoita}.
\newblock {Propagation of cosmic rays in heliosphere: The HELMOD model}.
\newblock {\em Advances in Space Research}, 62:2859--2879, November 2018.

\bibitem{2017ICRC...35...24V}
A.~{Vittino}, C.~{Evoli}, and D.~{Gaggero}.
\newblock {Cosmic-ray transport in the heliosphere with HELIOPROP}.
\newblock {\em International Cosmic Ray Conference}, 35:24, January 2017.

\bibitem{1958ApJ...128..664P}
E.~N. {Parker}.
\newblock {Dynamics of the Interplanetary Gas and Magnetic Fields.}
\newblock {\em \apj}, 128:664, November 1958.

\bibitem{1999ApJ...520..204G}
J.~{Giacalone} and J.~R. {Jokipii}.
\newblock {The Transport of Cosmic Rays across a Turbulent Magnetic Field}.
\newblock {\em \apj}, 520:204--214, July 1999.

\bibitem{1977ApJ...213..861J}
J.~R. {Jokipii}, E.~H. {Levy}, and W.~B. {Hubbard}.
\newblock {Effects of particle drift on cosmic-ray transport. I - General
  properties, application to solar modulation}.
\newblock {\em \apj}, 213:861--868, May 1977.

\bibitem{1989ApJ...339..501B}
R.~A. {Burger} and M.~S. {Potgieter}.
\newblock {The calculation of neutral sheet drift in two-dimensional cosmic-ray
  modulation models}.
\newblock {\em \apj}, 339:501--511, April 1989.

\bibitem{2003AdSpR..32..645F}
S.~E.~S. {Ferreira}, M.~S. {Potgieter}, and B.~{Heber}.
\newblock {Particle drift effects on cosmic ray modulation during solar
  maximum}.
\newblock {\em Advances in Space Research}, 32:645--650, August 2003.

\bibitem{2015ApJ...810..141P}
M.~S. {Potgieter}, E.~E. {Vos}, R.~{Munini}, M.~{Boezio}, and V.~{Di Felice}.
\newblock {Modulation of Galactic Electrons in the Heliosphere during the
  Unusual Solar Minimum of 2006-2009: A Modeling Approach}.
\newblock {\em \apj}, 810:141, September 2015.

\bibitem{1993ApJ...403..760P}
M.~S. {Potgieter}, J.~A. {Le Roux}, L.~F. {Burlaga}, and F.~B. {McDonald}.
\newblock {The role of merged interaction regions and drafts in the
  heliospheric modulation of cosmic rays beyond 20 AU - A computer simulation}.
\newblock {\em \apj}, 403:760--768, February 1993.

\bibitem{1993AdSpR..13..239P}
M.~S. {Potgieter}.
\newblock {Time-dependent cosmic-ray modulation - Role of drifts and
  interaction regions}.
\newblock {\em Advances in Space Research}, 13:239--249, June 1993.

\bibitem{2019ApJ...878....6L}
Xi~{Luo}, Marius~S. {Potgieter}, Veronica {Bindi}, Ming {Zhang}, and Xueshang
  {Feng}.
\newblock {A Numerical Study of Cosmic Proton Modulation Using AMS-02
  Observations}.
\newblock {\em \apj}, 878(1):6, Jun 2019.

\bibitem{2016A&A...591A..94G}
A.~{Ghelfi}, F.~{Barao}, L.~{Derome}, and D.~{Maurin}.
\newblock {Non-parametric determination of H and He interstellar fluxes from
  cosmic-ray data}.
\newblock {\em \aap}, 591:A94, June 2016.

\bibitem{2003ApJ...598L..63G}
N.~{Gopalswamy}, A.~{Lara}, S.~{Yashiro}, and R.~A. {Howard}.
\newblock {Coronal Mass Ejections and Solar Polarity Reversal}.
\newblock {\em \apjl}, 598:L63--L66, November 2003.

\bibitem{2014SoPh..289.3381K}
N.~{Karna}, S.~A. {Hess Webber}, and W.~D. {Pesnell}.
\newblock {Using Polar Coronal Hole Area Measurements to Determine the Solar
  Polar Magnetic Field Reversal in Solar Cycle 24}.
\newblock {\em \solphys}, 289:3381--3390, September 2014.

\bibitem{2015ApJ...798..114S}
X.~{Sun}, J.~T. {Hoeksema}, Y.~{Liu}, and J.~{Zhao}.
\newblock {On Polar Magnetic Field Reversal and Surface Flux Transport During
  Solar Cycle 24}.
\newblock {\em \apj}, 798:114, January 2015.

\bibitem{2015GeAe..55..969T}
A.~G. {Tlatov}, D.~V. {Dormidontov}, R.~V. {Kirpichev}, M.~P. {Pashchenko},
  A.~D. {Shramko}, V.~S. {Peshcherov}, V.~M. {Grigoryev}, M.~L. {Demidov}, and
  P.~M. {Svidskii}.
\newblock {Study of some characteristics of large-scale solar magnetic fields
  during the global field polarity reversal according to observations at the
  telescope-magnetograph Kislovodsk Observatory}.
\newblock {\em Geomagnetism and Aeronomy}, 55:969--975, December 2015.

\bibitem{2016KPCB...32...78P}
M.~I. {Pishkalo} and U.~M. {Leiko}.
\newblock {Dynamics of the circumpolar magnetic field of the Sun at a maximum
  of cycle 24}.
\newblock {\em Kinematics and Physics of Celestial Bodies}, 32:78--85, March
  2016.

\bibitem{2016ApJ...823L..15G}
N.~{Gopalswamy}, S.~{Yashiro}, and S.~{Akiyama}.
\newblock {Unusual Polar Conditions in Solar Cycle 24 and Their Implications
  for Cycle 25}.
\newblock {\em \apjl}, 823:L15, May 2016.

\bibitem{2018AA...618A.148J}
P.~{Janardhan}, K.~{Fujiki}, M.~{Ingale}, S.~K. {Bisoi}, and D.~{Rout}.
\newblock {Solar cycle 24: An unusual polar field reversal}.
\newblock {\em \aap}, 618:A148, October 2018.

\bibitem{2017AdSpR..60..833G}
A.~{Ghelfi}, D.~{Maurin}, A.~{Cheminet}, L.~{Derome}, G.~{Hubert}, and
  F.~{Melot}.
\newblock {Neutron monitors and muon detectors for solar modulation studies: 2.
  $\phi$ time series}.
\newblock {\em Advances in Space Research}, 60:833--847, August 2017.

\bibitem{2019JGRA..124.2367K}
Sergey~A. {Koldobskiy}, Veronica {Bindi}, Claudio {Corti}, Gennady~A.
  {Kovaltsov}, and Ilya~G. {Usoskin}.
\newblock {Validation of the Neutron Monitor Yield Function Using Data From
  AMS-02 Experiment, 2011-2017}.
\newblock {\em Journal of Geophysical Research (Space Physics)},
  124(4):2367--2379, Apr 2019.

\bibitem{2007ApJ...670.1149M}
J.~{Minnie}, J.~W. {Bieber}, W.~H. {Matthaeus}, and R.~A. {Burger}.
\newblock {Suppression of Particle Drifts by Turbulence}.
\newblock {\em \apj}, 670:1149--1158, December 2007.

\bibitem{2019EPJWC.20901004M}
N.~{Marcelli}, O.~{Adriani}, G.~C. {Barbarino}, G.~A. {Bazilevskaya},
  R.~{Bellotti}, M.~{Boezio}, E.~A. {Bogomolov}, M.~{Bongi}, V.~{Bonvicini},
  and S.~{Bottai}.
\newblock {Time dependence of the helium flux measured by PAMELA}.
\newblock In {\em European Physical Journal Web of Conferences}, volume 209 of
  {\em European Physical Journal Web of Conferences}, page 01004, Sep 2019.

\end{thebibliography}
\bibliographystyle{unsrt}

\newpage
\appendix

\section{The comparison of LIS in Paper I}\label{sec:liss}
\begin{figure}[!hbt]
\centering
   \includegraphics[width=.8\textwidth]{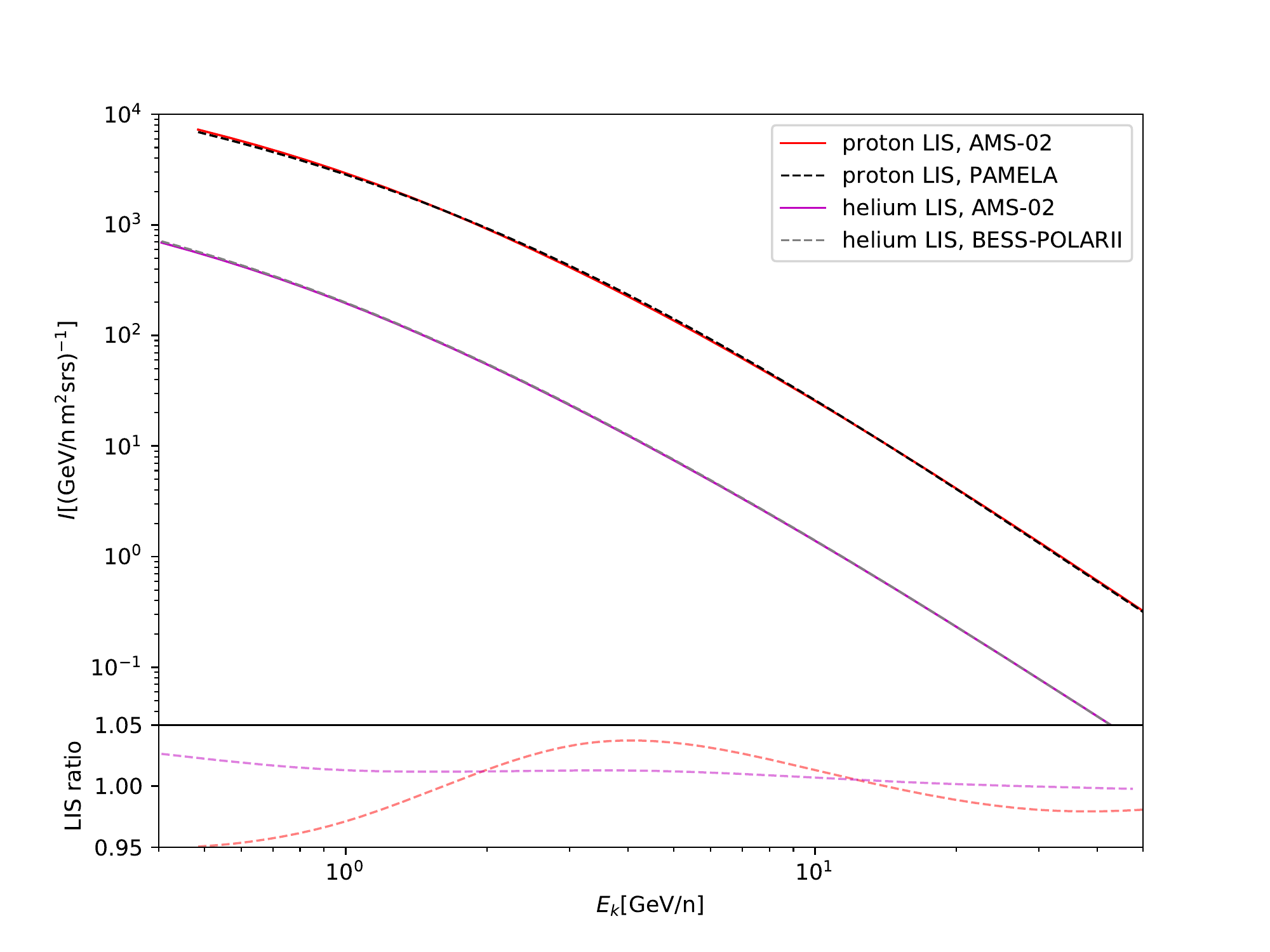}
   \caption{The comparison of LIS of proton and helium from the Paper I and this 
work.} \label{fg:liss}

\end{figure}

In the Paper I, the proton (helium) LIS is based on the Voyager 1 and PAMELA 
(BESS-POLARII) data.
We show the comparison of LIS obtained in this work and these in the Paper I at 
the top panel of Figure \ref{fg:liss}.
In the bottom panel, we show the ratio of LIS relative to these in the Paper I. 
The differences of the LIS are no more than $5\%$.

\section{Full drift in the positive polarity cycle}\label{sec:full}
\begin{figure}[!hbt]
\centering
   \includegraphics[width=.8\textwidth]{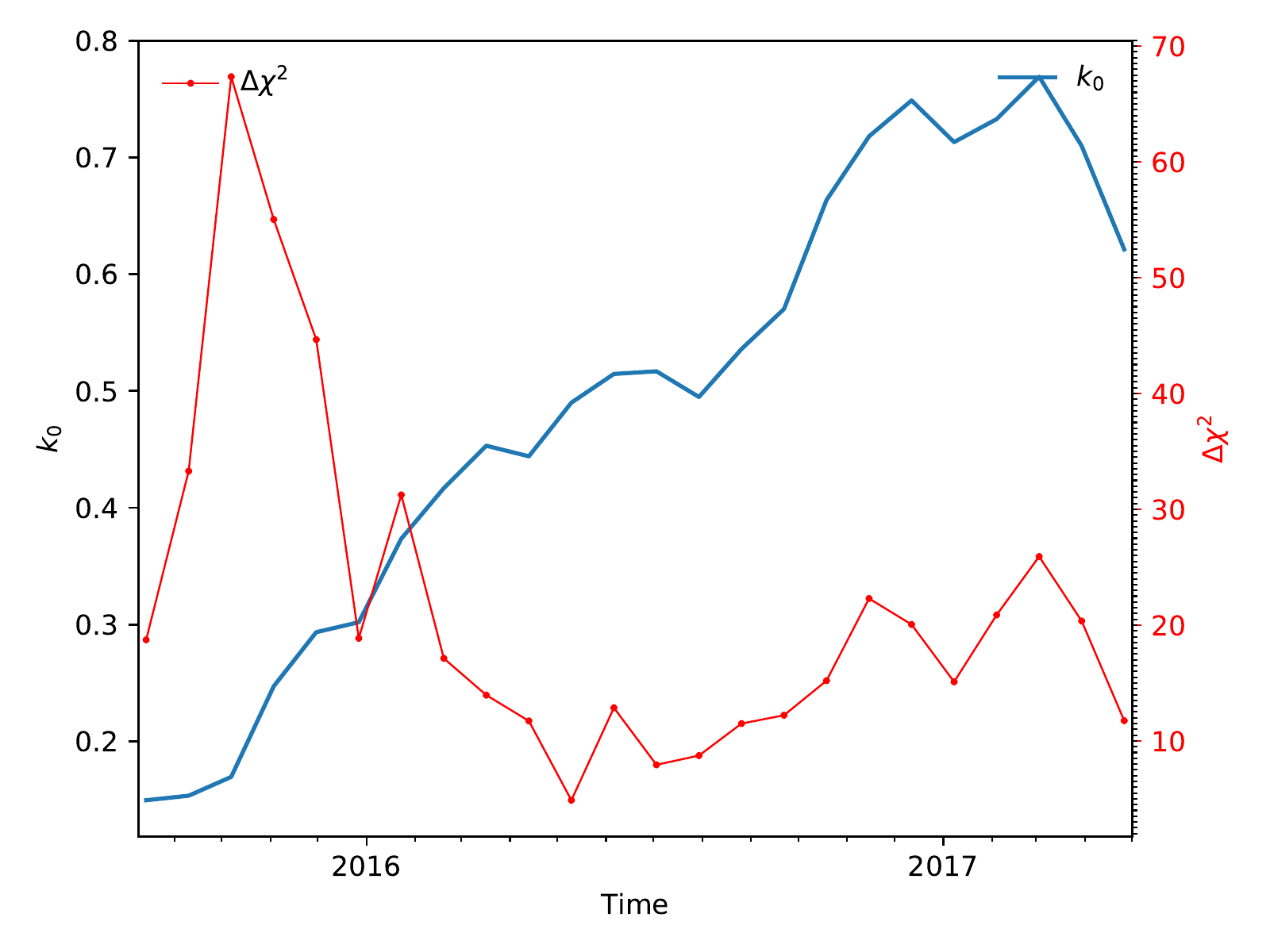}
   \caption{The time profile of $k_0$ with the assumption of the positive polarity and the full drift. The $\chi^2$ difference between the full drift scenario and the suppressed drift scenario is marked as $\Delta \chi^2$.} \label{fg:kd10}

\end{figure}

Under the assumption of the positive polarity and the full drift effect, the 
time profile of $k_0$
is shown in Figure \ref{fg:kd10}.
The difference of $\chi^2$ between the full drift and the suppressed drift
scenario are also shown.
The full drift scenario leads to very small $k_0$ and larger $\chi^2$.

%\newpage
\section{The degeneracy between $k_0$ and $k_d$}\label{sec:deg}
\begin{figure}[!hbt]
     \centering
     \includegraphics[width=0.8\textwidth]{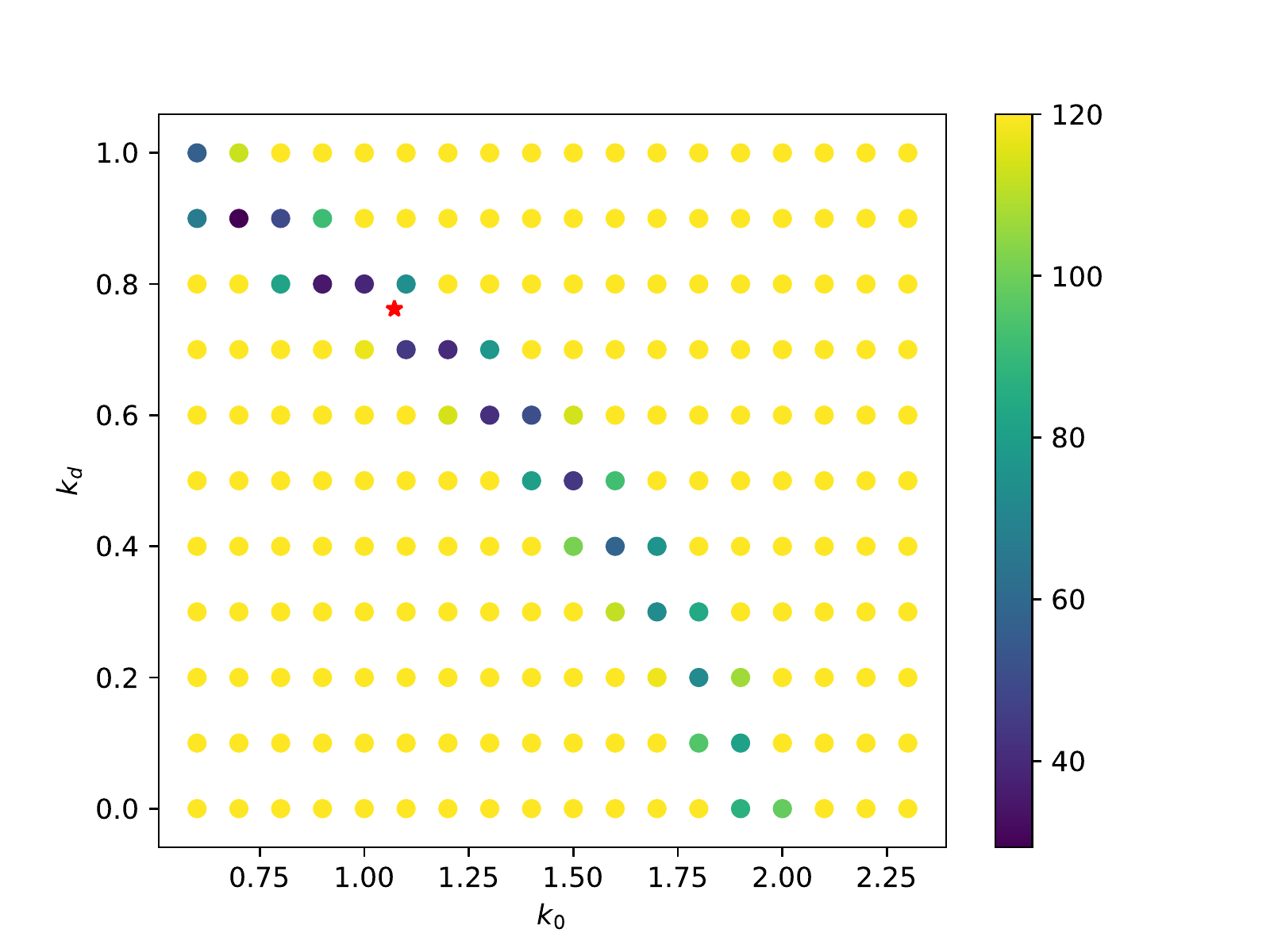}
     \caption{The degeneracy between $k_0$ and $k_d$. The red star indicates the best fit value. The color indicates the $\chi^2$ value in the grid.}
     \label{fg:k0kd}
\end{figure}

One example of the degeneracy between the diffusion and the drift is shown in 
Figure \ref{fg:k0kd}.
The computed spectra is compared to the proton data taken during 2017/04/13 to 2017/05/09.
$k_0$ and $k_d$ are taken as model input parameters and $\chi^2$ values are computed on the grid.
It can be seen that, there is a obvious degeneracy between $k_0$ and $k_d$.
Since the drift effect are opposite for proton and antiproton, the future monthly antiproton data
from AMS-02 is critical to break the degeneracy and improve our understanding on the modulation process.

\end{document}